\documentclass{article}
\usepackage{spconf,amsmath,graphicx,url,subfig, cite,booktabs}
\usepackage{amsfonts,amssymb}
\usepackage{xcolor}
\usepackage{makecell}
\usepackage[hidelinks]{hyperref}
\usepackage{svg}
\usepackage{multirow}
\usepackage{booktabs} 

\title{BS-PLCNet: Band-split Packet Loss Concealment Network with Multi-task Learning Framework and Multi-discriminators}
\vspace{-0.2cm} 
\name{Zihan Zhang$^{1,2}$, Jiayao Sun$^{1}$, Xianjun Xia$^2$, Chuanzeng Huang$^2$, Yijian Xiao$^{2}$, Lei Xie$^{1*}$\vspace{-0.5em}\thanks{$^*$: Corresponding author.}}
\address{
  $^1$Audio, Speech and Language Processing Group (ASLP@NPU), School of Computer Science, \\ Northwestern Polytechnical University, Xi'an, China\\
  $^2$ByteDance, China
  \vspace{-1em}
  }

\begin{document}
\vspace{-4pt}
\maketitle
%

\vspace{-2pt}
\begin{abstract}
Packet loss is a common and unavoidable problem in voice over internet phone (VoIP) systems. To deal with the problem, we propose a band-split packet loss concealment network (BS-PLCNet). Specifically, we split the full-band signal into wide-band (0-8kHz) and high-band (8-24kHz). The wide-band signals are processed by a gated convolutional recurrent network (GCRN), while the high-band counterpart is processed by a simple GRU network. To ensure high speech quality and automatic speech recognition (ASR) compatibility, multi-task learning (MTL) framework including fundamental frequency (f0) prediction, linguistic awareness, and multi-discriminators are used. The proposed approach tied for $1^{st}$ place in the ICASSP 2024 PLC Challenge.
\end{abstract}
\vspace{-2pt}
\begin{keywords}
Packet loss concealment, generative adversarial network, band split, multi-task learning
\end{keywords}
\vspace{-8pt}
\section{Introduction}
\vspace{-8pt}
\label{sec:intro}

In recent years, deep learning has shown great potential in packet loss concealment (PLC). 
However, long-gap packet losses and high-quality restoration still challenges the model performance.
As a flagship event of the ICASSP 2024 SP grand challenge, the audio deep PLC challenge is designed to solve these problems.
In this challenge, our team N\&B submitted a band-split PLC network (BS-PLCNet) with multi-task learning (MTL) framework and multi-discriminators. 
Due to the computational complexity of full-band signal processing and the fact that structured harmonics are mainly found in wide-band signals, we split the frequency band into 0-8kHz (wide-band) and 8-24kHz (high-band). To improve the naturalness of the generated audio and get better automatic speech recognition (ASR) compatibility, fundamental frequency (f0) prediction loss, and linguistic aware loss are adopted as auxiliary tasks along with PLC. Multi-discriminators, including multi-frequency discriminators (MFD), multi-period discriminators (MPD), and MetricGAN discriminators, are also used to improve speech quality.

\vspace{-6pt}
\section{Proposed method}
\vspace{-6pt}
\label{sec:method}

\subsection{BS-PLCNet generator}
\vspace{-6pt}
To better capture the dominant frequency and energy, and generate higher quality audio, we design our BS-PLCNet particularly in spectral-domain.
After analyzing different sub-band strategies, including pseudo quadrature mirror filter bank (PQMF), spectrum splitting, and band-split, we find that splitting the frequency band into two parts and processing them with different modules can yield the best performance. 
Specifically, the input complex-valued spectra from short-time Fourier transform (STFT) are compressed and stacked by real and imaginary parts, then divided into wide-band and high-band. 
Given that the harmonic structures are mainly found in the wide-band, and the human auditory system exhibits greater sensitivity to the wide-band, we use a large GCRN-based module to process the wide-band. For the high-band, we use a lightweight GRU-based module.

\begin{figure*}[t]
    \centering
    \includegraphics[scale=0.65]{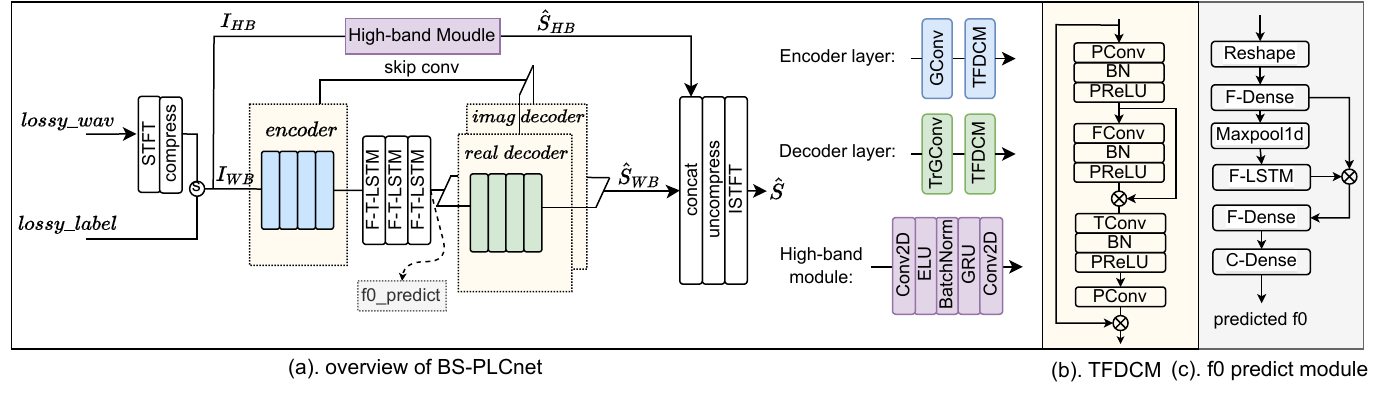}\vspace{-0.3cm}
    \caption{The proposed BS-PLCNet (a), the structure of the TFDCM module (b) and the structure of the f0 prediction module (c).}\vspace{-0.6cm}
    \label{fig:model}
\end{figure*}

The backbone of the wide-band module is based on our previous work~\cite{sun2023multi}. As shown in Fig.~\ref{fig:model}(a), the wide-band module consists of 4 encoder layers, an F-T-LSTM bottleneck layer, and 4 decoder layers. We also add an f0 prediction module to improve the signal coherence. 
The high-band module consists of a 2D convolution (Conv2d) layer, an ELU layer, a batchnorm layer, and a GRU layer followed by another Conv2d layer.

We use an MTL framework, where f0 prediction and linguistic awareness are adopted as auxiliary tasks along with PLC. The structure of the f0 prediction module is shown in Fig.~\ref{fig:model}(c). Same as~\cite{zhang2022multi}, we use the frequency feature-wise linear modulation (F-FiLM) as our f0 prediction module. 

To utilize linguistic awareness, we adopt Whisper-large\footnote{\url{https://huggingface.co/openai/whisper-large-v2}}, a ASR model trained on 680k hours of multilingual data. Specifically, we use the Whisper encoder to extract representations from the BS-PLCNet's output and the labels.
\vspace{-6pt}

\vspace{-8pt}
\subsection{Multi-discriminators}
\vspace{-6pt}
We combine the time-domain discriminator and time-frequency (T-F) domain discriminator to classify the original and the recovered audio. The time-domain multi-period discriminator (MPD) is from~\cite{kong2020hifi}, the sequence of sample points is folded into a 2D plane at a certain periodicity and is processed via Conv2D layers. The T-F domain multi-frequency discriminator (MFD)~\cite{tian2020tfgan} operates at different window lengths. In the MFD, sequential Conv2D modules are used for downsampling, discriminating at different spectral resolutions.  The MetricGAN+ discriminator~\cite{fu2021metricgan+} is also introduced to help the PLC model optimize perceptual metrics better.

\vspace{-8pt}
\subsection{Loss function}
\vspace{-6pt}
The speech loss is composed of a power-law compressed phase aware (PLCPA) loss~\cite{zhang2022multi} and time-domain MAE loss.
{
\setlength\abovedisplayskip{2.0pt}
\setlength\belowdisplayskip{2.0pt}
\begin{equation}
\mathcal{L}_\text{speech}=\mathcal{L}_\text{PLCPA} + \mathcal{L}_\text{MAE}.
\end{equation}
}
The f0 prediction loss $\mathcal{L}_\text{f0}$ and the linguistic aware loss $\mathcal{L}_\text{whisper}$ are all MAE feature loss.
The overall multi-loss for the generator is:
{
\setlength\abovedisplayskip{2.0pt}
\setlength\belowdisplayskip{2.0pt}
\begin{equation}
\mathcal{L}_\text{final}=\mathcal{L}_\text{PLCPA} + \mathcal{L}_\text{MAE} + \alpha \cdot \mathcal{L}_\text{f0} + \beta \cdot \mathcal{L}_\text{whisper} + \mathcal{L}_\text{GAN},
\end{equation}
}
where $\alpha$ equals 1e-1, $\beta$ equals 1e-3. $\mathcal{L}_\text{GAN}$ is the LSGAN loss and the MetricGAN+ loss for the generator.

\vspace{-6pt}
\section{EXPERIMENTS}
\vspace{-4pt}
\label{sec:EXPERIMENTS}

\vspace{-4pt}
\subsection{Datasets and experiments setup}
\vspace{-4pt}
For the training data, we use clean data from 5th DNS Challenge to simulate the lossy signals. Only English data was selected, totaling about 300 hours. Additional data from Librispeech and Interspeech 2022 PLC Challenge datasets are also used for the wide-band module. We use the Gilbert-Elliott model~\cite{mushkin1989capacity} to model the conditional consecutive packet loss. The packet loss rate is set to not exceed 50\%.

Window length and frame shift are 20 ms and 10 ms, respectively, and 960-point STFT is applied. The details of the wide-band module are the same as~\cite{sun2023multi}. For the high-band module, the first Conv2D has an output channel of 128, and the GRU layer has a hidden state of 128. Lastly, a gain decay operation is applied to the spectrum if the lost packet frame number exceeds 7. Our samples are available\footnote{\url{https://zzhdzdz.github.io/BS-PLCNet}}.

\vspace{-8pt}
\subsection{Results and conclusions}
\vspace{-6pt}
\begin{table}[]
\small
\scriptsize
\centering
 \caption{The objective scores on the blind test set.}
 \vspace{-8pt}
 \label{tab:corrcoef}
  \resizebox{\linewidth}{!}{
\begin{tabular}{@{}lcccc@{}}
\toprule
    & PLC MOS & DNS MOS  & WER(\%)  \\ \midrule
Lossy &  2.69  & 2.89   &  21.34       \\
BS-PLCNet & 3.88 & 3.34 & 16.75   \\ 
\quad + f0 predict & 3.96 &  3.36 & 16.21   \\ 
\quad + f0 + whisper loss & 3.93 & 3.34 & \textbf{15.58}  \\ 
\quad + f0 + whisper + MetricGAN & \textbf{ 4.02} & \textbf{3.39} & 15.71  \\
\bottomrule
\end{tabular}
}
\end{table}
\begin{table}[]
\small
\centering
\vspace{-3pt}
 \caption{The final scores on blind test set}
 \vspace{-8pt}
 \label{tab:results}
  \resizebox{\linewidth}{!}{
\begin{tabular}{@{}lccccccc@{}}
\toprule
    & P.804 Discontinuity  & P.804 Overall & Word Accuracy & Final Score \\ \midrule
Raw(lossy) &  2.47 & 2.37   & 0.83 & 0.51  \\ 
BS-PLCNet & \textbf{3.82} & \textbf{3.44} & \textbf{0.84} &\textbf{0.72} \\
\bottomrule
\end{tabular}
}
\vspace{-17pt}
\end{table}
To validate the model performance, speech MOS and WER\footnote{\url{https://github.com/wenet-e2e/wenet}} are calculated on the 2024 and 2022 blind test set, respectively. 
We have the following conclusions. The application of f0 prediction is beneficial with higher MOS and lower WER. While using Whisper loss did not significantly improve the speech MOS, it effectively reduce the WER. Last, MetricGAN also improves speech quality. Our submitted model achieved $1^{st}$ place (tied) in the challenge ranking. The number of parameters is 3.81M. Its RTF is 0.26, tested on Intel(R) Xeon(R) E5-2678 v3 @2.50GHz using a single thread (speed-up by ONNX).



\vspace{-6pt}
\bibliographystyle{IEEEbib}
\let\oldbibliography\thebibliography 
\renewcommand{\thebibliography}[1]{ 
  \oldbibliography{#1}
  \setlength{\itemsep}{-1pt} 
}
\bibliography{strings,refs}

\end{document}